\documentclass[apj,twocolumn]{emulateapj}

\usepackage{natbib, aas_macros}
\usepackage[normalem]{ulem}

\newcommand{\io}{i_{0}}
\newcommand{\gio}{(g-i)_{0}}
\newcommand{\ebv}{\rm{E(B-V)}}
\newcommand{\mm}{(m-M)_{0}}
\newcommand{\gmh}{\rm{[M/H]}}
\newcommand{\feh}{\rm{[Fe/H]}}

\newcommand{\rr}{\rm{R_{25}}}
\newcommand{\sqdeg}{\rm{deg^2}}

\shorttitle{Interacting Galaxies of the M81 Group}
\shortauthors{Okamoto et al.}

\begin{document}

\title{A Hyper Suprime-Cam View of \\the Interacting Galaxies of the M81 Group$^{\dag}$}

\author{Sakurako Okamoto\altaffilmark{1}}
\author{Nobuo Arimoto\altaffilmark{2,3}}
\author{Annette M.N. Ferguson\altaffilmark{4}}
\author{Edouard J. Bernard\altaffilmark{4}}
\author{Mike J. Irwin\altaffilmark{5}}
\author{Yoshihiko Yamada\altaffilmark{6}}
\author{Yousuke Utsumi\altaffilmark{7}}

\affil{$^{1}$Shanghai Astronomical Observatory, Nandan road, Shanghai, 200030, China\\
  $^{2}$Subaru Telescope, National Astronomical Observatory of Japan, 650 North A'ohoku Place Hilo, HI 96720, U.S.A.\\
  $^{3}$The Graduate University for Advanced Studies, Osawa 2-21-1, Mitaka, Tokyo, 181-8588, Japan\\
  $^{4}$Institute for Astronomy, University of Edinburgh, Royal Observatory, Blackford Hill, Edinburgh, EH9 3HJ U.K.\\
  $^{5}$Institute of Astronomy, University of Cambridge, Madingley Road, Cambridge CB3 0HA, U.K.\\
  $^{6}$National Astronomical Observatory of Japan, Osawa 2-21-1, Mitaka, Tokyo, 181-8588, Japan\\
  $^{7}$Hiroshima Astrophysical Science Center, Hiroshima University,
  Kagamiyama 1-3-1, Higashi-Hiroshima, Hiroshima 739-8526, Japan}

\altaffiltext{\dag}{Based on data collected at Subaru Telescope, which is operated by the National Astronomical Observatory of Japan.}

\begin{abstract}
  We present the first results of a wide-field mapping survey of the
  M81 group conducted with Hyper Suprime-Cam on the Subaru Telescope.
  Our deep photometry reaches $\sim2$ magnitudes below the tip of the
  red giant branch (RGB) and reveals the spatial distribution of both old
  and young stars over an area of $\sim 100\times115$ kpc at the
  distance of M81. The young stars ($\sim30-160$ Myr old) closely
  follow the neutral hydrogen distribution and can be found in a
  stellar stream between M81 and NGC\,3077 and in numerous 
  outlying stellar associations, including the known concentrations of Arp's Loop, 
  Holmberg\,IX, an arc in the halo of M82, BK3N, and the
  Garland. Many of these groupings do not have counterparts in the
  RGB maps, suggesting they may be genuinely young systems.  Our survey
  also reveals for the first time the very extended ($\geq 2\times\rr$) halos of 
  RGB stars around M81, M82 and NGC\,3077, as well as  faint tidal streams that 
  link these systems. The halos of M82 and NGC\,3077 exhibit
  highly disturbed morphologies, presumably a consequence of the recent gravitational
  encounter and their ongoing disruption.  While the halos of M81, NGC\,3077 and 
  the inner halo of M82 have the similar $\gio$ colors, the outer halo of M82 is 
  significantly bluer indicating it is more metal-poor.  
  Remarkably, our deep panoramic view of the M81 group demonstrates that 
  the complexity long-known to be present in HI is equally matched in the low
  surface brightness stellar component.   
  \end{abstract}


\section{Introduction}

Over the last decade, deep studies of nearby galaxies have led to the
discovery of vast stellar envelopes that are often rich in
substructure \citep{Mihos2005ApJ, Martinez2010AJ}.  
These components are naturally predicted in
models of hierarchical galaxy assembly and their observed properties
place important constraints on the amount, nature and history of
satellite accretion \citep[e.g.][]{Bullock2005ApJ,Pillepich2014MNRAS}.
Due to their very low surface brightness, one of the most effective
ways of mapping the peripheral regions of galaxies is through resolved
star studies.  For example, dedicated surveys of red giant branch
(RGB) stars around M31 have revealed a stellar
halo extending to more than $\sim200$~kpc that is dominated by tidal
debris features \citep[e.g.][]{Ibata2001Natur,Ferguson2002AJ,
 McConnachie2009Natur,Ibata2014ApJ}.  Similarily,
the Sloan Digital Sky Survey (SDSS) has been used to explore 
main-sequence (MS) turn off stars in the halo of the Milky Way, leading 
to many discoveries of new substructures, satellites and a refined characterization 
of halo and thick disk properties \citep[e.g.][]{Belokurov2007ApJ}.

Using wide-field cameras equipped to 8m class telescopes, it has
recently become possible to extend these studies to systems beyond
the Local Group \citep[e.g.][]{Mouhcine2010ApJ,Barker2012MNRAS,
Crnojevic2013MNRAS}.  Located at a distance of 3.6 Mpc 
\citep{Freedman1994ApJ}, M81, is a prime target for wide-field mapping 
of its resolved stellar content.  
Spectacular neutral hydrogen images have demonstrated the significant tidal
interactions between M81 and its two brightest neighbors, M82 and NGC\,3077, 
which modelling suggests have taken place in the last 300 Myrs
\citep[e.g.][]{vanderHulst1979AA, Yun1994Nature, Yun1999IAUS,
 Chynoweth2008AJ}.  Deep photometry from the Hubble Space Telescope
(HST) has been used to argue that the outlying HI concentrations
Arp's Loop (AL), and Holmberg\,IX
(HoIX) may be tidal dwarf galaxies formed as a result of these
interactions \citep{Makarova2002AA, deMello2008AJ, Sabbi2008ApJ}.  In
NGC\,3077, 90\% of the atomic hydrogen is located eastward of the
center, in the tidal arm called ``the Garland" where young stars have
been observed \citep{Karachentsev1985MNRAS, Sakai2001ApJ,
 Weisz2008ApJ}.  Several other young star concentrations have been
associated with peaks in the HI gas \citep{Durrell2004IAUS,
 Sun2005ApJ, Davidge2008PASP, Mouhcine2009MNRAS, Chiboucas2013AJ},
however, the global properties of this population throughout the M81
group are still poorly known.

The old stellar content around M81 has also been studied using large telescopes.  \citet{Barker2009AJ}
found the evidence for a faint, extended structural component beyond
the bright optical disk of M81 from wide-field images taken by
Subaru/Suprime-Cam.  They detected no colour gradient 
in this structure out to 44~kpc; \citet{Monachesi2013ApJ} used 
HST pointings to extend this result to 50~kpc.  \citet{Chiboucas2013AJ} 
confirmed 12 new dwarf satellites as members of the M81 group, discovered from 
a $65~\sqdeg$ survey with CFHT/MegaCam.

In this paper, we present the first results from a deep wide-field
imaging survey of the M81 group that we are conducting with the new
prime-focus imager, Hyper Suprime-Cam (HSC), on the Subaru Telescope.
We report on the analysis of the inner $4~\sqdeg$ area,
corresponding to a region spanning $100\times115$~kpc at the distance
of the galaxy, which reveals the first truly panoramic view of the low 
surface brightness stellar component. The observations and data reduction are described in
Section \ref{sec: obs}. Section \ref{sec: cmd} and \ref{sec: spatial} present our analysis
and results, which are discussed and conclusions drawn in Section
\ref{sec: discussion}.

For this paper, we adopt a distance modulus for M81 and associated
systems of $\mm=27.79$ \citep{Radburn2011ApJS}, position angles for
M81, M82, and NGC\,3077 of $157^{\circ}, 67.5^{\circ}$, and
$55.0^{\circ}$ east of north,  $\rr$ radii for M81, M82, NGC\,3077 and HoIX of $13.8\arcmin,
5.6\arcmin$, $2.7\arcmin$ and $2.5\arcmin$, respectively
\citep{deVaucouleurs1991rc3, Karachentsev2004AJ}.

\section{Observations and data reduction}
\label{sec: obs}

We observed the central region of the M81 group in the $g$- and
$i$-bands using 4 pointings of Subaru/HSC during the nights of January
21 and 22, 2015 (PI: S. Okamoto; Proposal ID: S14B-101) with the seeing 
ranged from 0.6\arcsec to 0.9\arcsec.  The HSC
consists of 104 CCD detectors and provides a field-of-view of 1.76
$\sqdeg$ with a pixel scale of 0.17\arcsec \citep{Miyazaki2012SPIE}.
The observations were obtained as part of a survey to map the M81
group with 7 HSC pointings.  In this paper, we focus on the inner
$4~\sqdeg$ of our survey which overlaps the SDSS footprint
\citep{York2000AJ}.  

The raw images were processed using the HSC pipeline (version 3.2.2),
which is based on a software suite being developed for the Large
Synoptic Survey Telescope (LSST) project \citep{Ivezic2008arXiv,
  Axelrod2010SPIE}.  For the processed images, the DAOPHOT in
IRAF was used to obtain the PSF photometry of resolved stars
\citep{Stetson1987PASP}.  Astrometric and photometric calibrations
were done using the SDSS catalog. Artificial star tests were performed
on some parts of the reduced images using the ADDSTAR in
DAOPHOT, and indicate that our photometry is at least 50$\%$ complete
to 26 mag in both bands, except for the inner regions of galaxies.  We
separate point sources from extended sources in the same manner as for
Suprime-Cam images in \citet{Okamoto2012ApJ}.  Full details of the
observations and data reduction will be presented in a forthcoming
paper.

\section{The color-magnitude diagrams}
\label{sec: cmd}

%%%%%%%%%%%%%%%%%%%%%
%%%% CMD of all area (cmd_all)
\begin{figure}[t]
 \plotone{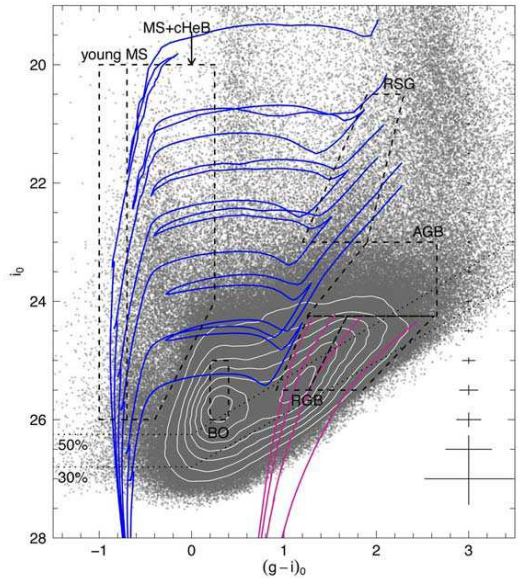}
 \caption{Dereddened CMD of stellar objects located within the central
   $4~\sqdeg$ area. The dashed boxes delineate the selection criteria
   for different stellar populations and are used to construct the
   maps presented in Figure \ref{fig: map}.  Theoretical Padova
   isochrones, adjusted at $\mm=27.79$, are shown for a 12 Gyr old
   population with $\gmh=-2.2,-1.75,-1.3,-0.75$ from the left to the
   right (magenta), and for a [M/H]=-0.75 population with ages of
   10,18,32,50,100,160 Myr from top to bottom (blue).  The dotted
   lines represent the completeness levels of $50\%$ and $30\%$.}
 \label{fig: CMD}
\end{figure}
%%%%
%%%%%%%%%%%%%%%%%%%%%

Figure \ref{fig: CMD} shows the resulting CMD of roughly 550,000 point
sources found in the whole $4~\sqdeg$ field.  The error bars represent
the photometric errors at $\gio=0$, as estimated by the artificial
star tests.  The Galactic extinction is taken from
\citet{Schlafly2011ApJ}.  Since the extinction varies across the
observed field,  
we apply a reddening correction to each source individually according to its location,
assuming a \citet{Fitzpatrick1999PASP} reddening law with
$\rm{R_V}$=3.1.  The central region within $r=15\arcmin$ of M82 in the reddening map shows significantly higher extinction $\ebv\sim0.16$ that includes the internal reddening of M82.  Therefore, we replace it with $\ebv=0.075$.

%%%%%%%%%%%%%%%%%%%%%
%%%% CMDs of galaxies (cmd_multi)
\begin{figure*}[t]
 \begin{center}
 \includegraphics[width=510pt]{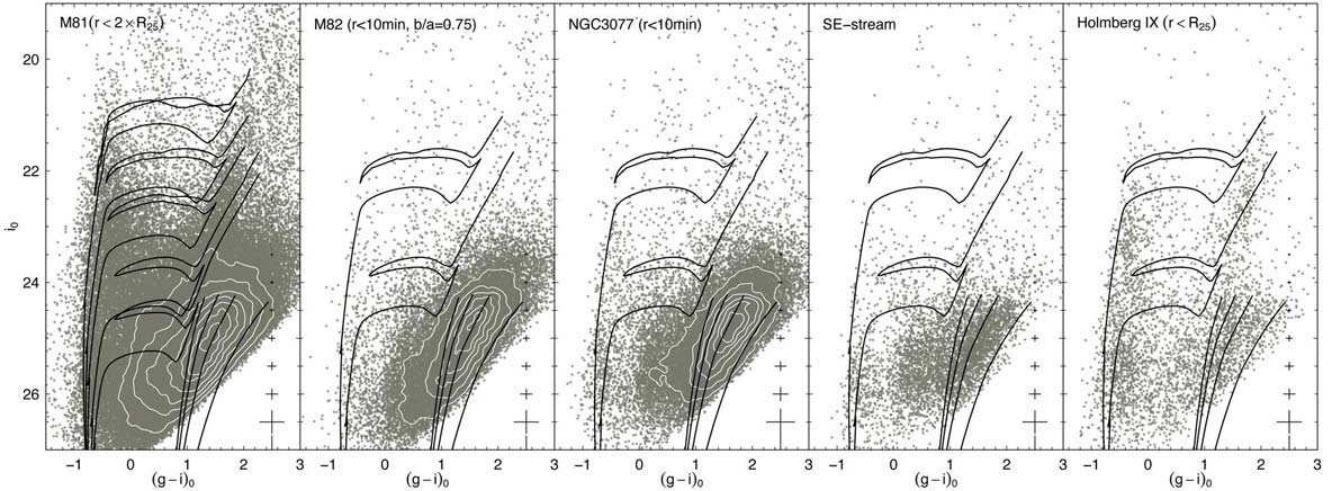} 
 \caption{Dereddened CMDs of stellar sources in the disk and halo (see text) of each galaxy, and in the SE-stream. The overplotted isochrones are the same as those in Figure \ref{fig: CMD}. }
  \label{fig: CMDs}
  \end{center}
\end{figure*}
%%%%
%%%%%%%%%%%%%%%%%%%%%

To aid in understanding the range of stellar populations present,
theoretical Padova isochrones are overlaid \citep{Bressan2012MNRAS}.
We find that tracks with metallicity $\gmh$ varying from $\sim$0.0 to
below $-1.0$ for young stars, and $\gmh\lesssim-1.0$ for old stars
provide a good description of the data, in agreement with other
studies of smaller regions \citep{Makarova2002AA, Williams2009AJ, Durrell2010ApJ,
  Kudritzki2012ApJ, Barker2009AJ, Monachesi2013ApJ}.  We use
$\gmh=-0.75$ as the fiducial value for the young population, and
overlay the 10-160 Myr old isochrones as blue solid
lines in Figure \ref{fig: CMD}.  For the old population, we plot
$\gmh=-2.2$ to $-0.75$ isochrones of 12 Gyr old as magenta
solid lines.

%%%%%%%%%%%%%%%%%%%%%
%%%% maps of all populations (map_multi)
\begin{figure*}[t]
 \begin{center}
 \includegraphics[width=510pt]{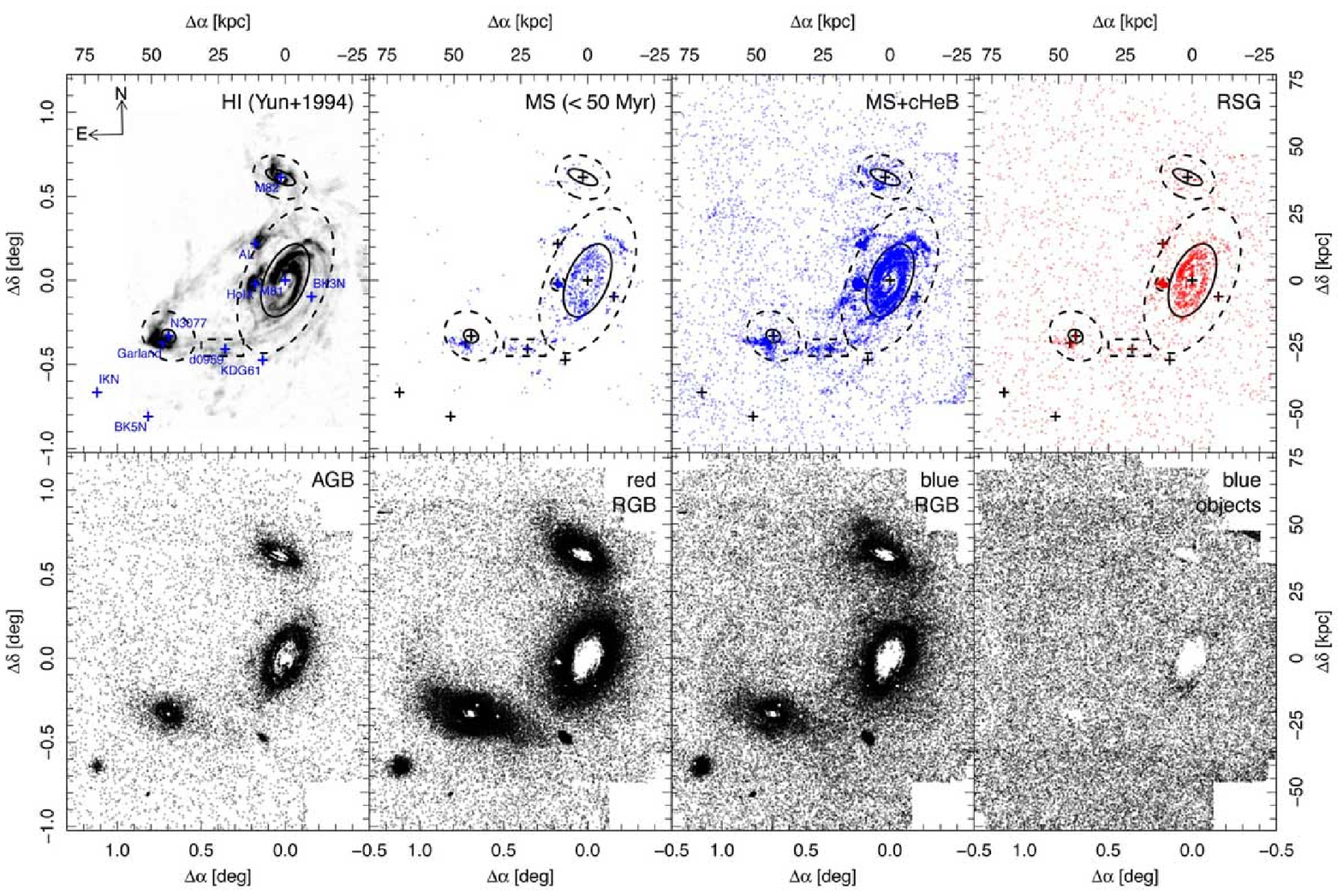} 
 \caption{The spatial distribution of stars in each evolutionary phase
   selected in Figure \ref{fig: CMD}.  Shown from the top left to the
   bottom right are the HI column density map taken from
   \citet{Yun1994Nature}, the spatial distributions of stars in evolutionary 
   phases and blue contaminants. The cross marks represent the centers of known M81
   group members. The solid lines are $\rr$ of galaxies, with axis ratios of 
   M82 and NGC\,3077 of $\rm{b/a}=0.38$ and $0.83$, and an inclination angle 
   of $i=58^{\circ}$ for M81, respectively.  The dashed lines outline 
   the regions used for the CMDs in Figure \ref{fig: CMDs}.}
  \label{fig: map}
  \end{center}
\end{figure*}
%%%%
%%%%%%%%%%%%%%%%%%%%%

Figure \ref{fig: CMD} is mostly populated by old RGB stars located at
$\io\ga24$ and $\gio\sim1.2$.  As discussed by
\citet{Barker2009AJ}, the over-density at $\io\sim26$ and
$\gio\sim0.3$ (labelled BO for `blue objects') mostly samples
unresolved background sources. On the blue side of the
foreground Galactic dwarf sequence (at $\gio\sim0.4$), young
MS and core Helium burning (cHeB) stars in the M81
group are found. We select stars in different evolutionary phases as
shown by the dashed boxes in Figure \ref{fig: CMD}: MS, cHeB, Red
Supergiant (RSG), Asymptotic Giant Branch (AGB) and RGB stars. The
boundaries were adopted to limit the number of foreground/background
contaminants. The young MS box mainly contains stars younger than
$\sim50$ Myr old while the MS$+$cHeB box is occupied by MS and post MS
stars of $<100$ Myr old.  On the red bright side, the polygon contains
RSGs about 25-160 Myr old with some contamination from Galactic disk stars.  Above the
RGB tip at $\io=24.25$, intermediate age ($\sim$ 0.5-8 Gyr) AGBs are
found \citep{Marigo2008AA}.  From the RGB tip to about 1.5 magnitude
below, the blue and red RGBs boxes contain stars older than 1 Gyr.  We
also examine the spatial distribution of BO sources. 
We note that the completeness of our photometry decreases towards the
red, as shown by the dashed lines, so our RGB sample may be biased
toward the bluer side.

Figure \ref{fig: CMDs} shows the dereddened CMDs of the disk and halo
of each galaxy, HoIX and the stream between M81 and NGC\,3077 (hereafter
the SE-stream). Stars within $r=2\times\rr$ (i.e.
$27.6\arcmin$ or 29~kpc) of M81, and within $10\arcmin$ (10.5~kpc) for M82 and NGC\,3077, within $r=\rr$ ($2.5\arcmin$ or 2.6~kpc) of HoIX are shown (see dashed lines in Figure \ref{fig: map}); an axis ratio of 0.75
has been used for M82 to take into account the varying flattening of
the stellar distribution in the outer regions.  We also select a
$15\arcmin \times 6\arcmin$ area for the SE-stream between M81 and NGC\,3077.  
Young blue MS and RSG can be easily seen in the M81 CMD, as well as
vast numbers of RGBs.  In M82 and NGC\,3077, RGBs are prominent and
some MSs exist, but few if any RSGs can be seen.  We note that the
area within $\rr$ of each galaxy could not be resolved 
due to crowding, so we miss the stars of the M81 and M82 disks, and in NGC\,3077 center. In the SE-stream, the most luminous MSs correspond to $\sim$32 Myr old.  
In HoIX, several MSs, cHeBs and RSGs exist.  Although we cannot resolve RGBs at the innermost ($<1\arcmin$) of HoIX, the number density of RGBs in HoIX are comparable to those of other regions at the same distance from M81.  Therefore, as discussed by \citet{Sabbi2008ApJ} with deeper HST images, most of the old component at HoIX may belong to M81 halo. 

\section{The spatial distributions of young and old components}
\label{sec: spatial}

Figure \ref{fig: map} shows the spatial distributions of HI gas
\citep{Yun1994Nature}, MS, cHeB, RSG, AGB, blue and red RGB stars, and
BOs defined in Figure \ref{fig: CMD}, without
correction for the completeness and 
contaminants.  In the upper panels, the solid ellipses represent the
$\rr$ radii of the three main galaxies. 

The distribution of the young populations traced by the young MS,
MS+cHeB and RSG stars agrees extremely well with that of the HI
distribution, except for the stream at the northwest of NGC\,3077 where
few stars are seen. The young MSs are mainly concentrated in the
spiral arms of M81, at the north-west side (hereafter NW-arm),
AL, HoIX, BK3N, the Garland, and SE-stream.  A
number of small clumps are also seen as they follow HI blobs around
these systems.  The stellar concentration in the NW-arm was identified
by \citet{Davidge2008PASP} and was suggested to be part of the M81 arm
by \citet{Barker2009AJ}. We confirm that it is connected to the
stellar concentration on the north arm.  The higher density regions in
the SE-stream have been reported either as clumps or as a dwarf galaxy
($\rm{d0959+68}$) in previous studies \citep{Durrell2004IAUS,
  Mouhcine2009MNRAS, Chiboucas2013AJ}.  As \citet{Chiboucas2013AJ}
discussed, these over-densities are clearly parts of a single stream.  Near
NGC\,3077, many MSs are found in the Garland and up to about 8~kpc to
the south and 10~kpc to the east where HI gas and dust emission have
been observed \citep{Walter2011ApJ}. In M82, a prominent stellar feature 
can be seen at $(\Delta\alpha,\Delta\delta)\sim(0.1,0.6)$, 
identified as an arc by \citet{Sun2005ApJ} and \citet{Davidge2008ApJ}. 
MSs are also distributed well beyond
the $\rr$ radius up to the projected distance of about 16~kpc from
M82.

%%%%%%%%%%%%%%%%%%%%%
%%%% contour of red RGB (map_rgbr_cont20sigma)
\begin{figure}[t]
 \plotone{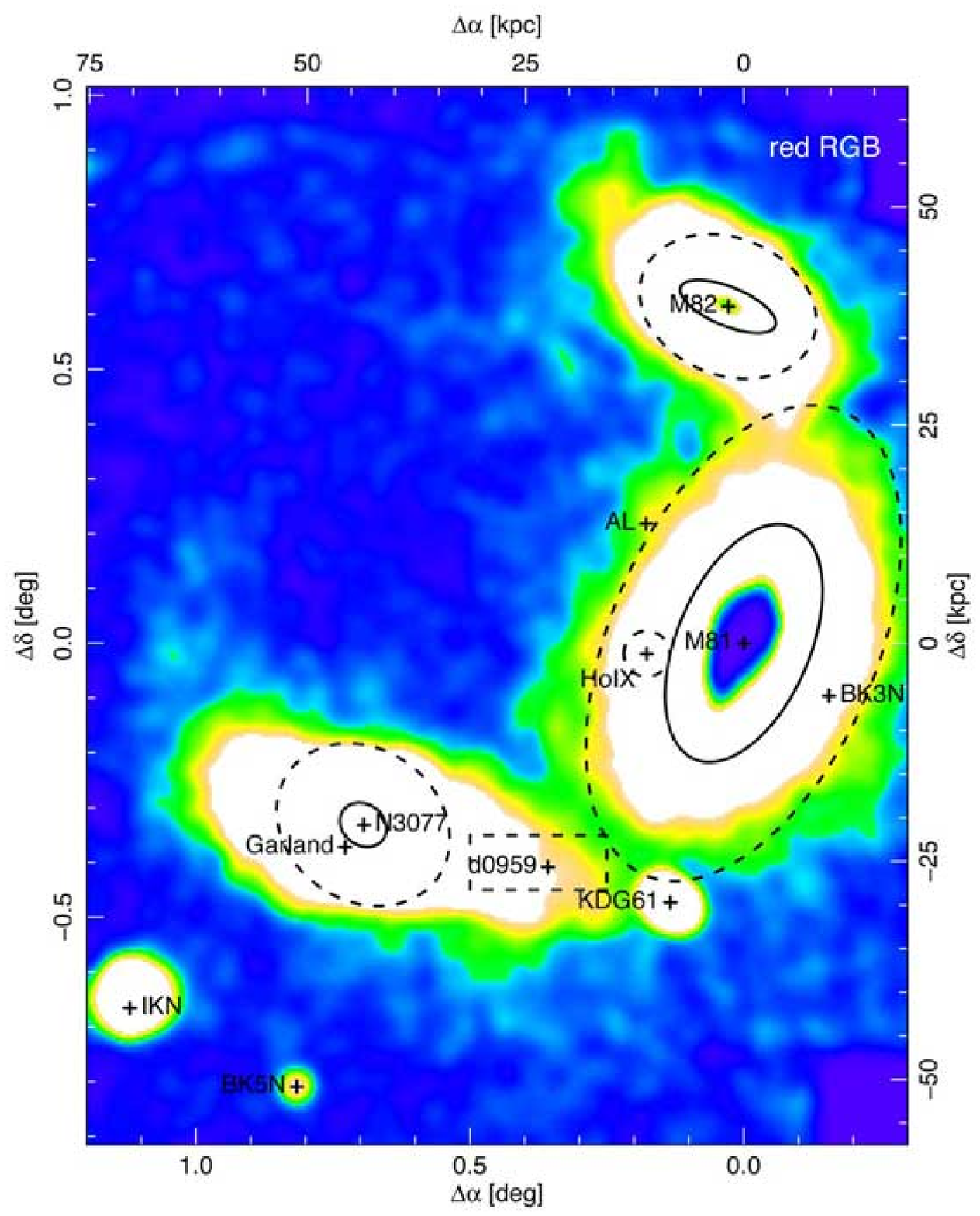}
 \caption{The isodensity contour map of red RGB stars, featured to faint structures 
 up to $20\sigma$ above the background level.  The kernel density is estimated with 
 the bandwidth of 1.2$\arcmin$. The marks and lines are the same as in Figure \ref{fig: map}.}
 \label{fig: map_cont}
\end{figure}
%%%%
%%%%%%%%%%%%%%%%%%%%%

\begin{figure*}[t]
 \begin{center}
 \includegraphics[width=510pt]{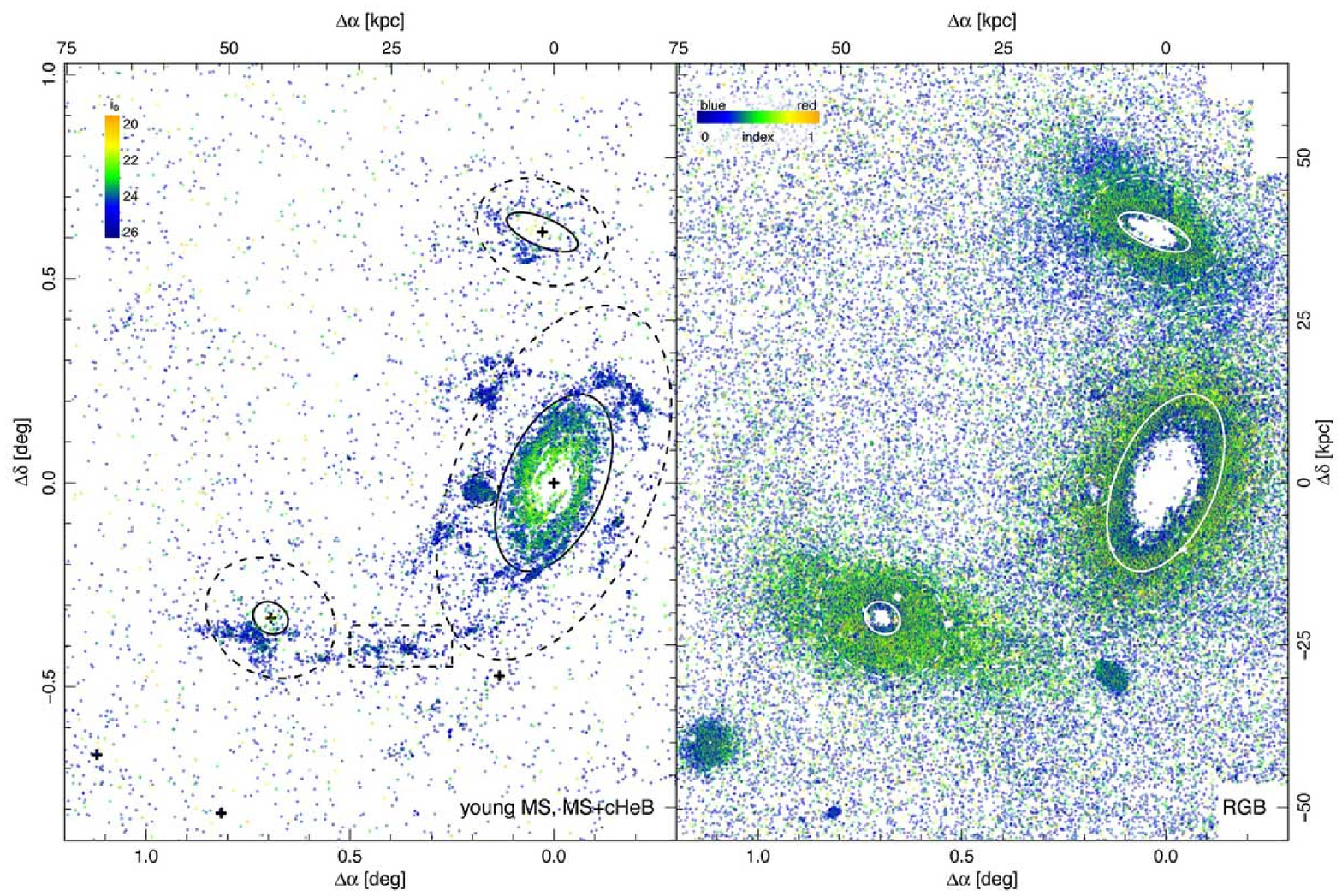} 
 \caption{$Left$: The spatial distribution of MS and cHeB stars that
   are color coded according to the luminosity with transparency.
   $Right$:The spatial distribution of RGB stars.  The color of each
   point represents the $\gio$ color of star with transparency.  The
   marks and lines are the same as in Figure \ref{fig: map}. }
  \label{fig: maps}
  \end{center}
\end{figure*}

Stars in the MS$+$cHeB box have a very similar distribution to that of
the young MSs.  The arc at the southeast of M82 appears to have a
clumpy shape, but it is an artificial appearance due to the overlap of
plotted points (see Figure \ref{fig: maps}). Interestingly, our maps
show that there is another young stellar feature on the opposite side
of M82 that appears to be aligned with the southern arc and may
therefore be related.  The selection box of MS$+$cHeB stars includes some
contamination, mainly from background blue objects, as can be seen by
the low-level uniform distribution of sources through out the area. In
the top-right panel of Figure \ref{fig: map}, the distribution of the
RSGs is almost the same as that of MSs and cHeBs. However, the fainter
substructures -- SE-stream, M82-arc, some clumps around HoIX and AL,
and BK3N -- can not be seen in this map due to the shorter lifetime
and the lower number of RSGs compared to MSs.

%%%%%%%%%%%%%%%%%%%%%
%%%% MS RGB map (map_multi_msrgb)
%%%%
%%%%%%%%%%%%%%%%%%%%%

In contrast, the older populations (AGB, RGBs) have a much smoother
distribution than the younger stars.  Old stars are mainly embedded in
the halos which reach far beyond the $\rr$ radii of three galaxies. The sizes of
these RGB halos are considerable, and they may even overlap (see Figure \ref{fig: map_cont}).  
While the M82 halo seems to be more extended in the blue RGB
map, the halos of M81 and NGC\,3077 are more prominent in the red RGB
map, suggesting they have a higher mean metallicity.  In the contour map of red RGBs in 
Figure \ref{fig: map_cont}, a tidal stream between M81 and M82 can clearly be seen, 
and the outer regions of M82 and NGC\,3077 exhibit an ``S-shaped" morphology.  The
dwarf galaxies IKN, BK5N, and KDG\,61 cannot be seen in the maps of
young stars, but appear as over-densities of old populations, implying
they have not formed as a result of the recent interaction.

In the bottom-right panel of Figure \ref{fig: map}, we plot the
distribution of BOs at $25<\io<26$ and $0.2<\gio<0.4$.
The uniform distribution of these sources supports their
identification as contaminants, since the star/galaxy separation in
our photometry degrades at magnitude fainter than about 24 mag in both
$g$- and $i$-bands.

\section{Discussion and Summary}
\label{sec: discussion}

We find that the young intra-group population in the M81 group
traces the filamentary structures of the HI gas connecting M81, M82,
and NGC\,3077, confirming the results of several smaller field-of-view
studies.  The left panel of Figure \ref{fig: maps} shows the
spatial distribution of stars in young MS and MS$+$cHeB boxes of
Figure \ref{fig: CMD}, which are color coded according to $i$-band
magnitude in a transparent manner, so that colors of overlapping points
represent the average color.  Bright stars are mainly located in the
inner disk of M81, while most of young stars in AL, NW-arm, BK3N,
Garland, and other debris features are fainter than $\io\sim24$ and
have similar luminosity distributions to the SE-stream, implying ages
of 30-160 Myr old.  This suggests that star formation in these tidal
features was synchronized, and may have stopped about 
30 Myrs ago. The SE-stream is slightly inclined from the
southeast to the northwest, whereas the tail of RGB stars in NGC\,3077
at the same location is extended toward the northeast-southwest. That
fact that these spatial distributions do not exactly match may
indicate that the SE-stream comes from material torn from M81 while
the RGB tail is material stripped from NGC\,3077.  The color of HoIX 
is slightly greener than other clumps and streams, meaning it includes 
some bright (younger) MS stars as we see in Figure \ref{fig: CMDs}.

The right panel of Figure \ref{fig: maps} shows the color distribution
of RGBs, which can be interpreted as a rough proxy for metallicity.
The bluest (index=0) and reddest (index=1) colors correspond to
$\gmh=-2.3$ and $-0.75$, respectively, assuming 12 Gyr old age.  
The RGBs between the solid and dashed lines of M81, M82 and
NGC\,3077 have the similar colors; the medians of the color indexes are
0.41, 0.40, 0.46 respectively, corresponding to $\gmh\sim-1.4$, $-1.4$ and $-1.3$.  
This M81 halo metallicity is slightly lower than the value of $\gmh\sim-1.1$ derived
from previous Subaru imagery and the estimation
$\feh=-1.2$ from deep HST photometry \citep{Barker2009AJ,Monachesi2013ApJ}.  
This might due to the missing metal-rich RGB stars in our photometry, since the
completeness gets worse at redder colors.  
The halo of M82 has a color gradient from the inner greener area to the outer
bluer region; the greener part is also extended toward the northeast
and along the direction of the young stellar arc.  Note that we
do not correct for the internal extinction of each galaxy, so it is
not possible from this study alone to determine if these are the bona
fide features of the M82 halo.  The tidal stream between M81 and M82
is also predominantly blue in color, indicating that this is material being
stripped from M82 onto M81.

In Figure \ref{fig: map_cont} and \ref{fig: maps}, the NGC\,3077 halo is extended far beyond
the $\rr$, and has a rhombus shape stretched in the east-west
direction. In the outermost region, an S-shape distortion can
be discerned, which is similar to what was found around M33
\citep{McConnachie2010ApJ}.  The component in the northwest appears to
reach a maximum projected radius from NGC\,3077 of $\sim$ 65~kpc, but
does not appear to trace the HI distribution in this region.  The S-shaped
structure is typical of an interacting dwarf galaxy with a larger
companion \citep[e.g.][]{Pearrubia2009ApJ}.  Numerical modelling
suggests the encounters between NGC\,3077, M81 and M82 took place
$\sim$ 200-300 Myr ago \citep{Yun1999IAUS}, which may not leave enough
time to restore equilibrium in the NGC\,3077 halo.  We will return to
the topic of the stellar streams in the M81 group in a later paper.

The close encounters between M81, M82 and NGC\,3077 induced star
formation in tidally stripped gas.  As a consequence, new stellar
concentrations were born out of these HI rich clumps, many of which
lie far from the main bodies of the galaxies.  Of these
concentrations, only AL appears to have a clear counterpart in the RGB
map. The presence of an older stellar component suggests that this
object, like the dwarf galaxies IKN, BK5N, and KDG 61, may not have a tidal origin.  The gravitational interactions between the M81 group
galaxies have also significantly perturbed their older stellar
components leading to disturbed halo morphologies and giant stellar streams
which appear to connect all three systems.  When combined
with our forthcoming HSC observations of the west side of
M81, these data will allow us to determine the true extent and nature of the intra-group
debris and map the halos of the M81 group galaxies to unprecedented distances.

\acknowledgments We are grateful to the entire staff at Subaru
Telescope and the HSC team.  We acknowledge the importance of
Maunakea within the indigenous Hawaiian community.
This paper makes use of software
developed for the LSST.  We thank the LSST Project for making their
code available as free software at http://dm.lsstcorp.org.  SO
acknowledges support from the CAS PIFI scheme.  AMNF and EJB acknowledge 
support from an STFC Consolidated Grant. This work was supported
by the grants of CAS (XDB09010100), NSFC (11333003), and JSPS (Grant-in-Aid for Young Scientists B, 26800103).

\end{document}